\documentclass{emulateapj}

\newcommand{\deltE}{\Delta\kern-1ptE}

\def\ion#1#2{#1\,{\sc #2}}

\input epsf

\usepackage{longtable}
\usepackage{amssymb}
\usepackage{graphicx}

\slugcomment{To be submitted to ApJ}
\shorttitle{High Energy 3D Photoionization Code}
\shortauthors{Ercolano et al.}
\begin{document}
\title{X-ray enabled MOCASSIN: a 3D code for photoionized media}
\author{Barbara Ercolano}
\affil{Harvard-Smithsonian Center for Astrophysics, MS-67
\\ 60 Garden Street, \\ Cambridge, MA 02138, USA}
\author{Peter R. Young}
\affil{STFC Rutherford Appleton Laboratory, \\Chilton, Didcot, \\
  Oxfordshire, OX11 0QX, UK}
\author{Jeremy J. Drake \& John C. Raymond}
\affil{Harvard-Smithsonian Center for Astrophysics\\ 
60 Garden Street, \\ Cambridge, MA 02138, USA}

\begin{abstract}
We present a new version of the fully 3D
photoionization and dust radiative transfer code, {\sc mocassin}, that uses a Monte Carlo
approach for the transfer of radiation. The X-ray enabled {\sc mocassin} allows a fully
geometry independent description of low-density gaseous
environments strongly photoionized by a radiation field extending from
radio to gamma rays.  The code has been thoroughly benchmarked against
other established codes routinely used in the literature, using simple
plane parallel models designed to test performance under standard
conditions. We show the results of our benchmarking exercise and
discuss applicability and limitations of the new code, which should be of guidance for
future astrophysical studies with {\sc mocassin}.  


\end{abstract}


\section{Introduction}\label{s:intro}

Photoionized environments characterize a wide range of astrophysical
problems involving sources of X--radiation. With the advent of new
technology used for instruments on board of (e.g.) {\it XMM-Newton} 
and {\it Chandra}, high resolution spectroscopy of such
environments has become a reality. Paerels et al. (2000), for instance,
observed the photoionized wind in Cygnus X--3 with {\it Chandra}
High Energy Transmission Grating Spectrometer (HETGS) showing the
discrete emission to be excited by recombination in a tenuous
X--ray--photoionized medium which is not symmetric with the source of
the wind. Other examples include 
the detection of several recombination emission lines (from
Fe~{\sc xxvi} at 1.78~{\AA} to N~{\sc vi} at 29.08~{\AA}), by
Jimenez-Garate et al. (2005) in a 50~ks
observation of the bright X-ray binary Hercules X-1 with {\it
  Chandra} HETGS. We also note the {\it Chandra} ACIS observations
of the deeply eclipsing cataclysmic variable DQ Herculis by Mukai et
al. (2003), who were able to pin down the origin of the soft X-rays 
from this system as being due to scattering of the unseen
central X-ray source, probably in an accretion disk wind.

 A number of 1D photoionization codes, including G.~Ferland's {\sc cloudy}
(Ferland et al. 1998) and T.~Kallman's {\sc xstar} (Kallman \&
 McCray 1980, Kallman \& Bautista 2001) continue to represent powerful analytical tools for
the analysis of astrophysical spectra from the X-ray to the infrared
regime. These codes are designed for diffuse, optically thin media, which may
also be irradiated by a non-thermal X-ray continuum, and have
been, for instance, applied to the modelling of Narrow Line Regions (NLRs) of Active
Galactic Nuclei (AGNs). Several other codes have been developed which specialize in emission
and reflection spectra from optically thick hot photoionized media,
irradiated by a non-thermal continuum extending to the hard X-ray
region, such as X--ray irradiated accretion disks (e.g. Ross \& Fabian, 
1993; Nayakshin et al. 2000; Dumont et al. 2000). 

To date, these and most other photoionization codes have numerically
solved the equations of radiative transfer (RT) under the assumption
of spherical symmetry or in plane parallel geometries, whereupon the problem is reduced
to a 1D calculation.  While very few real X-ray sources are
spherically symmetric, this approximation has been driven by the
available computing power and the complexity of the multi-dimensional
case. Mauche et al. (2004) developed a Monte Carlo code to investigate
the radiation transfer of Ly$\alpha$, He$\alpha$, and recombination
continua photons of H- and He-like C, N, O and Ne produced in the
atmosphere of a relativistic black hole accretion disk. This code,
however, while accounting for Compton scattering and photoabsorption
followed by recombination, does not calculate the ionization state of
the plasma. To our knowledge, there are currently no general,
self-consistent and publicly available X-ray photoionization and dust RT codes capable of
working in 3D. Although the 1D codes mentioned above are powerful
tools for the analysis of the pan-chromatic spectra of numerous
astrophysical environments, their application is restricted to
rather simplified geometries. 

The computational demand of realistic 3D simulations has recently come
within the reach of low-cost clusters.  Taking advantage of this,
the first self-consistent, 3D photoionization and dust RT code was
developed for the IR-UV regime using Monte Carlo
techniques (Ercolano et al. 2003a, 2005). The code, {\sc mocassin} 
(MOnte CArlo SimulationS of Ionized Nebulae)
was designed to build realistic models of photoionized environments
of arbitrary geometry and density distributions, and can
simultaneously treat the dust RT. The code can also treat
illumination from multiple point- or arbitrarily extended
sources. The fully parallel version of {\sc mocassin} (documented and
publicly available) is well-tested for
classical nebulae, according to standard photoionization benchmarks
(P\'equignot et al., 2001), and has been successfully applied to the
modeling of H~{\sc ii} regions (e.g. Ercolano, Bastian \&
Stasi\'{n}ska, 2007) and planetary nebulae (e.g. Ercolano et
  al. 2003b,c, 2004; Gon\c{c}alves et al., 2006; Schwarz \& Monteiro,
  2006; Wright, Ercolano \& Barlow, 2006). We now present a significantly 
improved version of the {\sc mocassin} code (version 3.00) 
which extends to the X-ray regime. In the
tradition of previous releases of the {\sc mocassin} code, the
X-ray version will also be made publicly available to the scientific
community shortly after publication of this article\footnote{Previous
  versions are available upon request from BE.}.

In Section~\ref{s:code} we
summarize the physical processes and atomic data added/modified
 in the new implementation
and discuss the applicability and limitations of the code in its
current form. In Section~\ref{s:bench} we compare our code to
established 1D codes for benchmark tests, comprising emission line
spectra from model NLR and from three thin low-density slab models 
illuminated by a hard continuum. A brief summary is given in Section~\ref{s:summary}.

\section{The X-ray enabled {\sc mocassin} code}\label{s:code}

\subsection{Basic philosophy and underlying assumptions}
The original {\sc mocassin} code 
was developed in order to provide a 3D modelling tool capable of
dealing with asymmetric and density and/or chemically inhomogeneous media, as well as, 
if required, multiple, non-centrally located, point and extended sources of exciting
radiation. The code can self-consistently treat the transfer of
ionizing and non-ionizing radiation through a medium composed of gas
and/or dust. 
The numerical techniques employed and the physical processes considered
by the code are described in detail by Ercolano et al. (2003a,
2005). In brief, {\sc mocassin} locally simulates the processes of absorption,
re-emission and scattering of photons as they diffuse through a
gaseous/dusty medium. The radiation field is expressed in terms of
 energy packets which are the calculation `quanta'. The energy packets
 are created at the illuminating source(s), which may be placed
 anywhere inside the grid and be of point-like or diffuse nature. Their trajectories through the nebula are
 computed, as the packets suffer scatterings, absorptions and
 re-emissions, according to the local
gas and dust opacities and emissivities. The packets' trajectories yield
a measure of the local radiation field, from which the local photoionization and 
recombination rates, as well as the heating and cooling integrals, are
determined.

It is worth reminding the reader of the physical assumptions that both
the previous and current versions of the {\sc mocassin} code rely
upon, which also define its applicability and limitations. 
\begin{itemize}
\item All physical processes affecting the gas are in steady
  state. This implies that the atomic physics and heating and cooling
  timescales are short 
compared to those of gas-motion or to the rate of change of the
  ionizing field. 
\item In some cases large optical depths may occur in the core of
  emission lines of astrophysically abundant ions. As described by
  Ercolano et al. (2005), resonant scattering is accounted for via an
  escape probability method. The underlying assumption is that
  photons in the line centre will be scattered many times in a region
  close to where they were originally emitted, until they
  finally escape through the line wings or are destroyed through
  continuum photo-absorption. While escape probability methods are
  often used in photoionization models, it is well known that these
  may lower the accuracy of calculations in high density environments
  (e.g. Avrett \& Loeser, 1988; Dumont et al. 2003). Furthermore, our current
  scheme assumes a static gas distribution, which would need to be modified to
 treat winds or accretion disks with very large shear.
\item The contribution from recombination is included for
  H--like, He--like and the lower Fe~{\sc xxiv} lines. Recombination,
  however, can dominate for several other lines in some extreme cases;
  in general, this process strengthens the 3s Fe L-lines compared
  to the 3d (Liedahl et al., 1990). Nevertheless, in the most
  commonly encountered cases, the contributions from recombination of
  He-like and H-like species are strongest.
\item Very high density environments are problematic and solutions may
  carry larger uncertainties. Apart from the line transfer
  issue mentioned above, a higher density
  limit of $\sim10^{13}$~cm$^{-3}$, at temperatures of $\sim$ 10$^4$~K, is imposed by our approximate treatment of three-body
  recombination and collisional ionization, which become important at
  higher densities (see also Ferland 2006). However, for highly
  ionized, hotter gas, e.g.  $\sim10^6$~K gas in the corona of an X-ray binary, the
  high density limit is increased to $\sim10^{15}$~cm$^{-3}$.
\item A treatment for unresolved transition arrays (UTAs), observed in several
spectra of AGNs (e.g. Sako at al. 2001), is not
currently included in {\sc mocassin}. Calculations were
presented by Behar, Sako and Kahn (2001) for inner shell n=2-3
photoexcitation of the 16 iron charge states Fe~{\sc i} through Fe~{\sc
xvi} and their data will be included in a future version.

\end{itemize}

\subsection{X-ray extension}

\subsubsection{Atomic Data}

{\sc mocassin}'s atomic database was updated in order to include the
latest data releases. Details of {\sc mocassin's} complete
database for the original version are given in Appendix~A of Ercolano
et al. (2003a). Data updates (additions and
replacements) implemented for version 3.00 are as follows:
\begin{itemize}
\item Free-bound emission for H~{\sc i},
  He~{\sc ii} and He~{\sc i} uses the data recently calculated by Ercolano \& Storey (2006).

\item The radiative and dielectronic rates of Badnell et al. (2003), Badnell (2006a,b),
Zatsarinny et al. (2003, 2004a,b, 2006), Colgan et al. (2003,
2004), Altun et al. (2004, 2006), Mitnik \& Badnell (2004) have been
included and are used as default. The data cover all elements up to
Zn including sequences up to Na-like electron target, all other
species are treated with the previously available data, except for
species included in the Nahar (1997,1999) data set, if chosen by the
user (see following item).

\item The total (radiative and dielectronic) rates of Nahar
  (1997, 1999) for recombination to all ions of carbon, nitrogen and oxygen and of Nahar (2000) for
  recombination to Si~{\sc i}, S~{\sc iii}, Ar~{\sc v}, Ca~{\sc vii}
  and Fe~{\sc xii} can now be included, if explicitly chosen. As well
  as the recombination data set mentioned as the default in the
  previous item, all data used in previous versions of {\sc mocassin} are still available for use and compatible with the X-ray version. 

\item The fits of Verner (1996) to Opacity Project (Seaton et al. 1993) data for the photoionization
cross-sections from all shells are used (the previous
versions used the same fitting procedures, but only allowed
photoionization from the outer valence electron shell).

\item The data set for collision strengths, transition probabilities and energy
levels has been substantially updated using version 5.2 of the {\sc chianti}
atomic database (Landi et al. 2006, Dere et al., 1997). The updates are
described in more detail in Section~\ref{s:chianti}. Further updates
are planned for the near future when version 6.0 of the database
becomes available. 

\item Recombination lines from hydrogenic species are calculated using
  fits to the line emissivities of Storey \& Hummer (1995). The data includes
  species up to Z=8, and temperatures extending to 10$^5$~K for
  Z\,$\geq$\,2. Species with Z\,$\geq$\,8 are calculated by scaling
  the He~{\sc ii} data. The same hydrogenic scaling to He~{\sc ii} is also
  used for species with 2\,$\leq$\,Z\,$\leq$\,8 when the local
  electron temperatures exceed the highest tabulated temperature of
  10$^5$~K.

\item Recombination lines from He-like species are calculated using
  fits to recombination data that will become publicly available
  in the next version of {\sc chianti} database. 
  The available total (direct+cascade) temperature-dependent
  recombination coefficients to the individual levels are used in the
  statistical equilibrium matrix that calculates the level
  populations, such that the final emissivities automatically include
  contributions from recombination. As well as the radiative-only
  rates derived from the Mewe et al. (1985) calculations for He-like
  iron, the dataset also includes more recent data of
  Porquet \& Daubau (2000), which include dielectronic contributions,
  when significant, for He-like ions with Z~=~6, 7, 8, 10, 12 and 14.

\item Recombination contributions to Li-like Fe~{\sc xxiv} transitions
  are calculated in a similar fashion to He-like contributions (see
  item above), using the temperature-dependent recombination
  coefficients of Gu et al. (2003). The data, however, extend only to
  the 3d level, transitions originating from higher levels (e.g. Fe~{\sc
  xxiv}~7.99~{\AA}) are calculated via hydrogenic scaling of the coefficients given in
  Storey \& Hummer (1995).

\item The ionization rate coefficients of atoms and ions by
electron impact now uses the fits presented by Voronov (1997). 
The collisional ionization and heating effects of the suprathermal secondary electrons
following inner-shell photoionization are computed using
approximations from Shull \& van Steenberg  (1985). Ionization by
suprathermal secondaries is generally only important for very low
ionization fractions. 
\end{itemize}

 \subsubsection{Inner-shell photoionization}
Previous versions of the {\sc mocassin} code were designed to
treat the transfer of low-energy ionizing radiation, typical of H~{\sc
  ii} regions and PNe. In these environments, only ionization from the
outer valence electron shell needs to be considered. We have now
lifted this limitation by including inner shell photoionization using
the Auger yields of Kaastra \& Mewe (1993) and the photoionization
cross-sections given by Verner et al. (1996). The solution of the
ionization balance equations is rendered more complicated by the
coupling of non-adjacent states through the emission of multiple
electrons by high energy photons. We adopt a similar iteration scheme
as that described by Ferland (2006) to solve the resulting ionization
balance matrix. 

\subsubsection{Thomson/Compton scattering}

The Compton cross-section is calculated via the Klein-Nishina
formula, which simplifies to the Thompson case at non-relativistic
energies. Compton heating and cooling contributions are calculated using standard formulae
(e.g. Rybicki \& Lightman, 1986). Bound Compton ionization and heating
from high energy photons ($\gtrsim$\,2.3 keV for hydrogen) are also taken into
account, using the formalism described by Ferland (2006, Eqn 294). 

The great advantage of a 3D Monte Carlo transfer technique is that
scattering events can be treated taking into account the real geometry
of the object. The re-processed directions of Compton-scattered
packets, for example, are calculated stochastically using
probability density functions based on the redistribution functions
obtained from the Klein-Nishina formulae. This is of particular
interest for the calculation of fluorescence spectra, as 
described in Drake, Ercolano \& Swartz (2007), Drake \& Ercolano
(2007a) and Drake \& Ercolano (2007b), but it can in principle be applied to any
line or continuum energy packet. 

\subsection{Energy levels, collision strengths and radiative decay
  rates}
\label{s:chianti}

Energy levels, collision strengths and radiative decay rates have been
updated with 
data from version 5.2 of the {\sc chianti} atomic database (Landi et al.,
2006; Dere et al., 1997). Since highly excited levels are generally not significantly
populated in photoionized plasmas, not all data from {\sc chianti} were
converted into the {\sc mocassin} format. Thus, for example, while the
{\sc chianti} models for many of the boron-like ions consist of 125 levels,
only data for the lowest 10 levels are currently used by {\sc
  mocassin}. The number of levels used for each iso-electronic sequence are
listed in Table~\ref{tbl.levels}. The
Maxwellian-averaged collision strengths, $\Upsilon$, are stored as
spline fits in {\sc chianti} \citep{dere97} which allow $\Upsilon$ to be
calculated for any temperature. For {\sc mocassin}, $\Upsilon$s
are calculated from the {\sc chianti} spline fits for the temperature range $2.0\le
\log\,T\le 6.4$ at 0.2~dex intervals.

\begin{figure}
\plotone{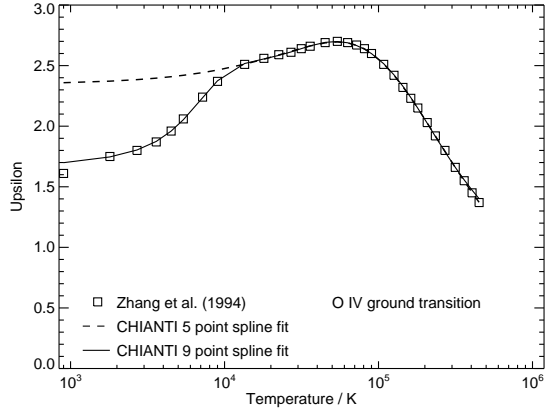}
\caption{\footnotesize Plot demonstrating the difference between five and nine point
  spline fits for the \ion{O}{iv}  $^2P_{1/2}$ -- $^2P_{3/2}$ ground
  transition. The original collision strength data are from
  \citet{zhang94}. The five
  point spline fit was optimized to the temperatures relevant to an
  electron ionized plasma ($\sim 10^5$~K) and it fails to reproduce
  the low-temperature region. The nine point spline gives a good fit to the low
  temperature data points needed for photoionized plasmas.
\label{fig.chianti-o4}}
\end{figure}

Up until version 4 of {\sc chianti} all collision data were fitted with a five
point spline, following the method of \citet{burgess92}. For some
transitions, however, a five point spline was not able to accurately fit
the entire set of collision strengths and thus it was necessary to
omit some of the original data points for the fit. The choice of which
data points to omit was influenced by the major application of
{\sc chianti} --
the analysis of emission line spectra from \emph{collisionally-ionized}
plasmas such as the solar corona -- thus care was taken to ensure the
fits were accurate in the temperature range where the ion was most
abundant. \citep[Tables of these temperature ranges are given in][for example.]{mazzotta98}
 Oftentimes this led to the collision
strength data at low temperatures being omitted from the fit. With
version 4 of {\sc chianti} \citep{young03} nine point spline fits to the
collision strength became available as an option, thus allowing
improved fits at low temperatures. As it is the low temperatures that
are important for \emph{photoionized} plasmas, the early {\sc chianti}
data-sets have been re-assessed to ensure that the low temperature data
points are being accurately reproduced. Poor fits were improved by
re-fitting the original data with nine point splines. An example of an improved fit to the
\ion{O}{iv} ground transition is shown in Fig.~\ref{fig.chianti-o4}.

In the process of adapting the {\sc chianti} data files for {\sc mocassin}, the
existing data within {\sc mocassin} have been compared with those in
{\sc chianti}. In general, {\sc mocassin} used the same or older
atomic data; in the latter case {\sc chianti} data simply replaced the
{\sc mocassin} data (after the checks described in the previous paragraph had been
performed). In some cases, {\sc mocassin} made use of better atomic data
(either more recent calculations, or data more suited to photoionized
plasmas), and so these data were assessed and added to {\sc chianti}.
The comparison between  {\sc chianti} and {\sc mocassin} also proved valuable for
identifying errors in the data files and a number of corrections have
been made to {\sc chianti}. These updates will appear in the next version of that
database.

\section{Benchmark Tests} \label{s:bench}

The complexity of the calculations involved in large scale
photoionization codes, where a number of coupled microphysical process
are at play, imposes the need for careful testing before predictions
from such codes may be trusted. In terms of energetics, care should be
taken that all ionization and recombination, heating and cooling
channels are correctly represented. 

The new code presented here has been thoroughly tested and compared, for the 1D case,
to the solutions to a number of benchmark models obtained by
independent codes routinely used in the literature. 
The set of benchmarks used was presented by P\'{e}quignot et al. (2001)
and comprises a plane-parallel simulation for conditions typical of
NLRs of AGNs and three low-density thin slab models irradiated by 
a strong broken power-law continuum radiation field, as described in
Table~10 of P\'{e}quignot et al. (2001), and reproduced here for
convenience (see Table~\ref{t:bpl}). For the NLR model 
the illuminating spectrum is a power law in energy units with slope
1.3, and the
ionization parameter, $U_{13.6}$~=~0.01 is defined as
$F_{13.6}$\,/\,($c\,\times\,n_{\rm H}$), where $F_{13.6}$ is the
incident photon flux above 13.6~eV. 
The ionization parameters of the three X-ray slab models, X0.1, X1.0
and X10.0 are $U_{100}$~=~0.1, 1.0 and 10.0, respectively, where the
definition of $U_{100}$ is analogous to that of $U_{13.6}$ above. 

The plane-parallel geometry assumed by the 1D test, can be mimicked
with our 3D code by modelling a thin, elongated cuboidal grid with
plane parallel diffuse illumination
coming at normal incidence from one of the smaller sides; the energy packets are only
allowed to escape from the side opposed to the illuminating side and all other sides act
as mirrors (for a discussion of the {\it mirror} technique see
Ercolano et al., 2003b).  The main input parameters for the benchmark tests
are given in Table~1 of P\'{e}quignot et al. (2001) and repeated in
our Table~\ref{t:benchinput} for convenience. 

We run a number of tests to estimate the random error due to Monte
Carlo sampling in order to establish an efficient, yet reliable,
combination of cell numbers and energy packets needed to sample them. We
adopted grids of 300 depth points and three by
three in the short axes, with the number of packets needed to achieve
convergence varying from one to eight million. Errors of the order of
0.5~\% in computed line strengths are commmon with this setup, with larger errors ($<$6\%)
sometimes obtained for the less abundant ions. 

We note that radiative transfer effects are largely unimportant for
the lines listed in the X-ray slab models. Furthermore, the models are
optically thin to ionizing radiation and electron temperature and
density, and ionization structure variations through the slabs 
are minimal. 

Before discussing the benchmark results in detail, we note that many
atomic data updates and code
developments have occurred in the past seven years (see Ercolano
2005). In particular,
the radiative and dielectronic
recombination coefficients calculated by Badnell et al. (2003), Badnell (2006a,b),
Zatsarinny et al. (2003, 2004a,b, 2006), Colgan et al. (2003,
2004), Altun et al. (2004, 2006), Mitnik \& Badnell (2004),
covering all elements up to Zn (plus Kr, Mo and Xe), including sequences
up to Na-like electron target\footnote{The data set is publicly available form
http://amdpp.phys.strath.ac.uk/tamoc/DATA/.}, represent a significant
improvement over earlier sets. 

\subsection{Narrow Line Region}

 We have organized the benchmark comparison in a similar fashion to
 P\'{e}quignot et al. (2001); for the NLR model we check 
that {\sc mocassin}'s prediction of each quantity falls within the
range of the values obtained by the other codes, listed in Table~10 of
P\'{e}quignot et al. (2001). In those cases where {\sc mocassin}'s
prediction lies outside the range, we compute the {\it isolation
  factors}, defined as the ratio of the value given by {\sc mocassin}
to the upper limit of the range, if our value is above the range, or
the ratio of the lower limit of the range to the value given by {\sc
  mocassin}, if our value is below the range.
Table~\ref{t:nlr} lists line fluxes, temperature and ionized helium
fraction as predicted by {\sc mocassin} (column 4) as well as the
minimum and maximum values (columns 2 and 3) predicted by the other
codes and given in Table~7 of P\'{e}quignot et al (2001). The 
isolation factors computed for {\sc mocassin} are given in the last
column of our Table~\ref{t:nlr}, where we have added a minus or plus
sign to indicate
whether our prediction lies below or above the range predicted by the
other codes.  

In P\'{e}quignot et al. (2001) it is stated that isolation factors
larger that 1.3 should be considered ``indicative of a very
significant departure and possible problem''.  
From Table~\ref{t:nlr} we find that only six
out of 48 quantities predicted carried an isolation factor with
absolute value larger that
1.3, with only one of these being larger that 2.0. We judge this to be
indicative of a high degree of consistency between the results of {\sc
  mocassin} and other codes, especially in the light of
the atomic physics updates that have occurred between the year 2001 and
today. Furthermore, the temperature and ionization structure of our
model is also in good agreement with those calculated by other codes
indicating that the differences in some line predictions can
probably be ascribed to atomic physics changes. 

The largest isolation factors were obtained for sulphur
ions.  The values in Table~\ref{t:nlr} show that our predictions for the [S~{\sc iii}]
lines are generally lower than those predicted by other
codes, and we suspect that dielectronic recombination effects may be at play in this
case. Rates for third row elements are poorly
known. In the case of recombination of S~{\sc iv} into S~{\sc
  iii} the rates calculated by Nahar (2000) are available. These
calculations used an ab-initio method to derive total (electron+ion)
recombination (Nahar \& Pradhan 1994, 1995) which enables the consideration of radiative and dielectronic
recombination processes in a unified manner. However not all modelers
choose to include
the total recombination coefficients due to the uncertainties inherent to the method
that relies on theoretical predictions of the resonance
positions. Here we have used the data of Colgan et al. (2003), which
were not available to the 2001 benchmark modelers.
Furthermore, no data was available in 2001 for S~{\sc
  iii}~$\rightarrow$~S~{\sc ii} dielectronic recombination and estimates were
used instead (e.g. Ali et al., 1991). Here we used the data of  Colgan
et al. (2004). It seems likely that some of the
scatter recorded for these lines is indeed due the different atomic
data available and assumptions made. 

\begin{figure}
\epsscale{1.0}
\plotone{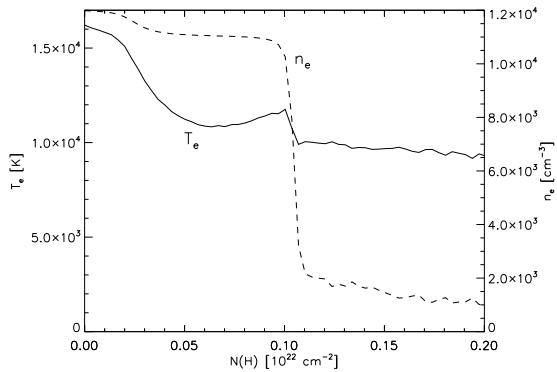}
\caption{\footnotesize NLR benchmark; electron temperature, $T_{\rm e}
  [K]$, and  density, $n_{\rm e} [cm^{-3}]$, as a function of hydrogen
  column density, $N_{\rm H} [cm^{-2}]$, through the slab. 
\label{f:tene}
}
\end{figure}

As a final note we should add that the temperature at the inner edge
calculated by {\sc mocassin} is slightly lower than the values
obtained by the other codes. This is to be expected since {\sc
  mocassin} uses a exact treatment of the radiative transfer,
including the diffuse component, while comparison codes approximate
the transfer of the diffuse component by either iterating along only
one direction or by adopting an ``outward-only'' approximation, but
along several directions. As noticed by P\'{e}quignot et al. (2001),
the kinetic temperature tends to be lower 
in the innermost layers of models with exact transfer, as the ionizing
radiation field there is softer. Figure~\ref{f:tene} shows electron
temperatures, $T_{\rm e} [K]$, and densities, $n_{\rm e} [cm^{-3}]$,
calculated with {\sc mocassin} as a function of column density in the slab. 

\subsection{X-ray irradiated plane-parallel slabs}

The P\'{e}quignot et al. (2001) benchmarks constitute
the first attempt to assess
the accuracy of photoionization codes in the X-ray regime. As well as
atomic data issues, already pointed out in the previous section, a 
further complication for this set of models is posed by the fact
that, in general, for the
 far-UV lines listed, the original benchmarks considered sums of
 multiplet lines in a rather liberal sense, so that {\it energetics}
 were privileged at the expense of accuracy of wavelengths (with the
 exceptions of some fine-structure optical lines). It is therefore possible
that part of the scatter amongst the original benchmark values
 arose because of the different way multiplets have been handled in
 different model calculations (P\'equignot, priv. comm.). This problem persists and should be kept in mind when
 looking at the comparisons performed in this
 article. The original benchmark tests were quite
 rough and their main 
 aim was to highlight gross discrepancies, particularly in derived temperatures.
 A new benchmarking exercise has yet to appear for the high-energy regime;
however it is still useful for us to check our code against the 2001
benchmarks as major problems in the thermal and ionization balances
would be uncovered by such a comparison. Furthermore, we propose that
the solutions presented here be taken as an improved set, obtained
with up-to-date atomic data.

 Models of  X-ray slabs are assessed by comparing, for each quantity,
 {\sc mocassin}'s prediction with the minimum and maximum limits of
 the range of values obtained by the other codes and listed in
 Tables~11-13 of P\'{e}quignot et al. (2001).

\begin{figure}
\epsscale{1.0}
\plotone{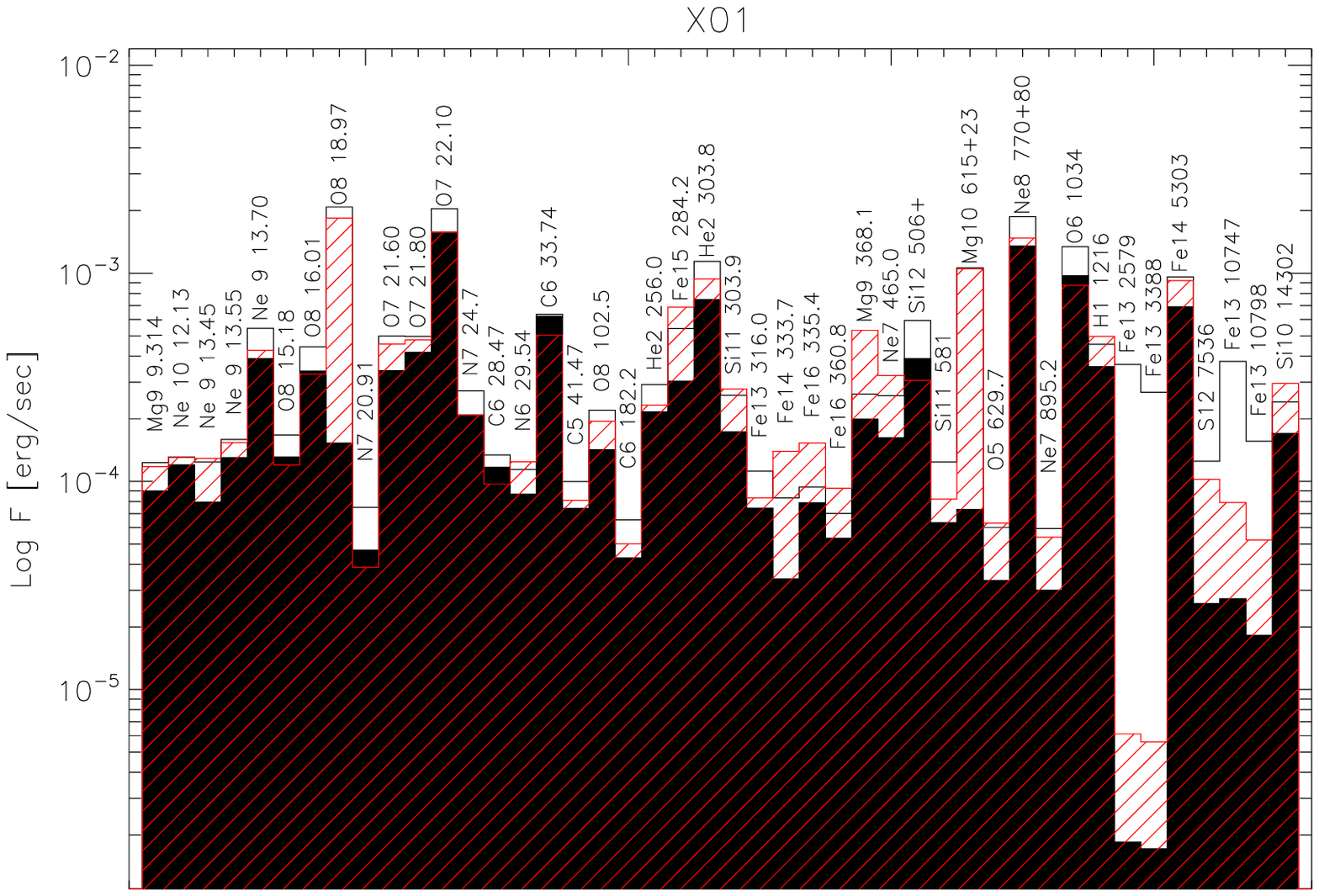}
\caption{\footnotesize X-ray slab; U$_{\rm X}$=0.1 (X01). The red,
  striped histograms illustrate {\sc mocassin's} results, while the
  filled and empty histograms represent the lower and higher limits of the
  range of predictions obtained by the codes in P\'equignot et al. (2001) . 
\label{f:x01}
}
\end{figure}

\begin{figure}
\epsscale{1.0}
\plotone{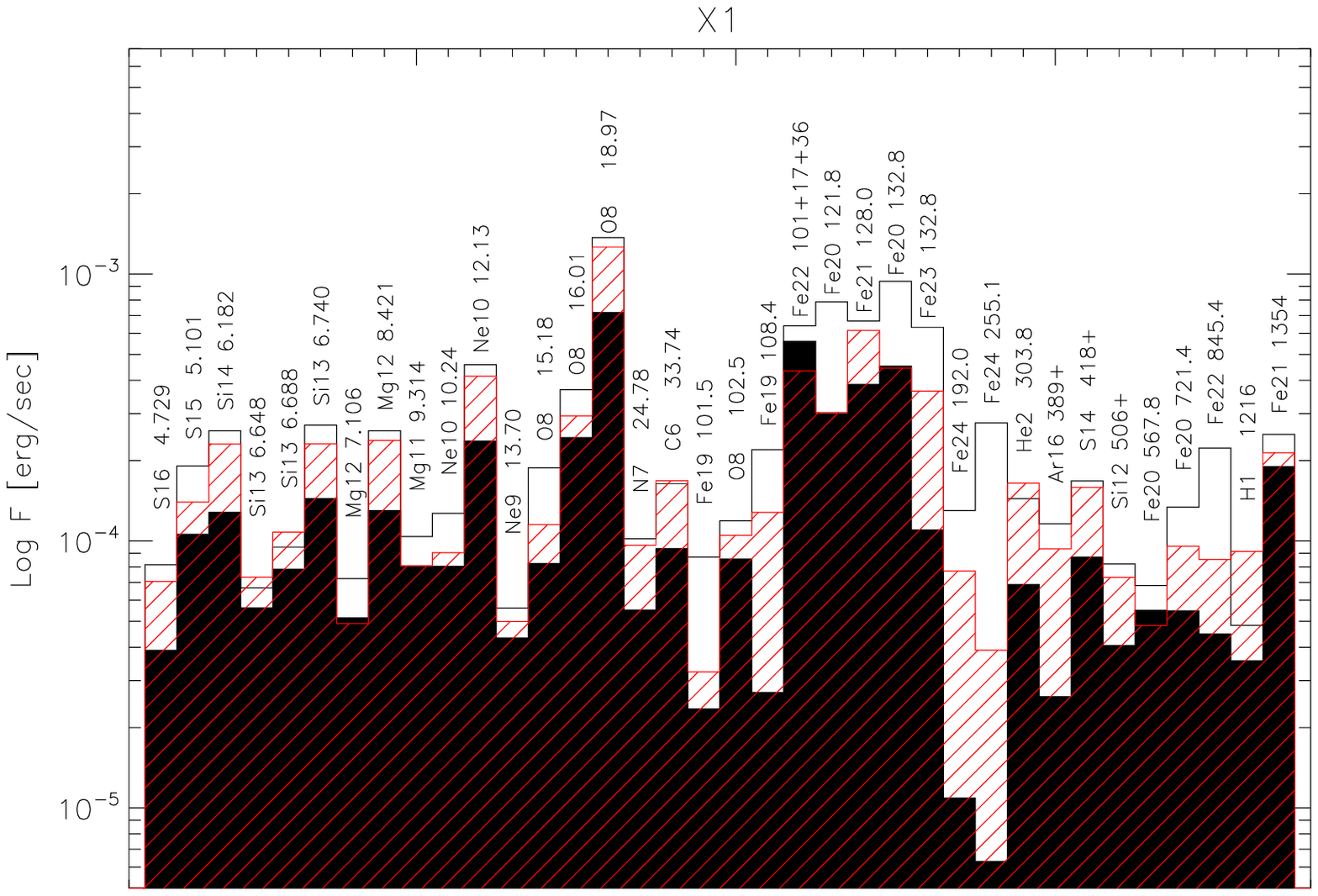}
\caption{\footnotesize X-ray slab; U$_{\rm X}$=1 (X1). The red,
  striped histograms illustrate {\sc mocassin's} results, while the
  filled and empty histograms represent the lower and higher limits of the
  range of predictions obtained by the codes in P\'equignot et al. (2001). 
\label{f:x1}
}
\end{figure}

\begin{figure}
\epsscale{1.0}
\plotone{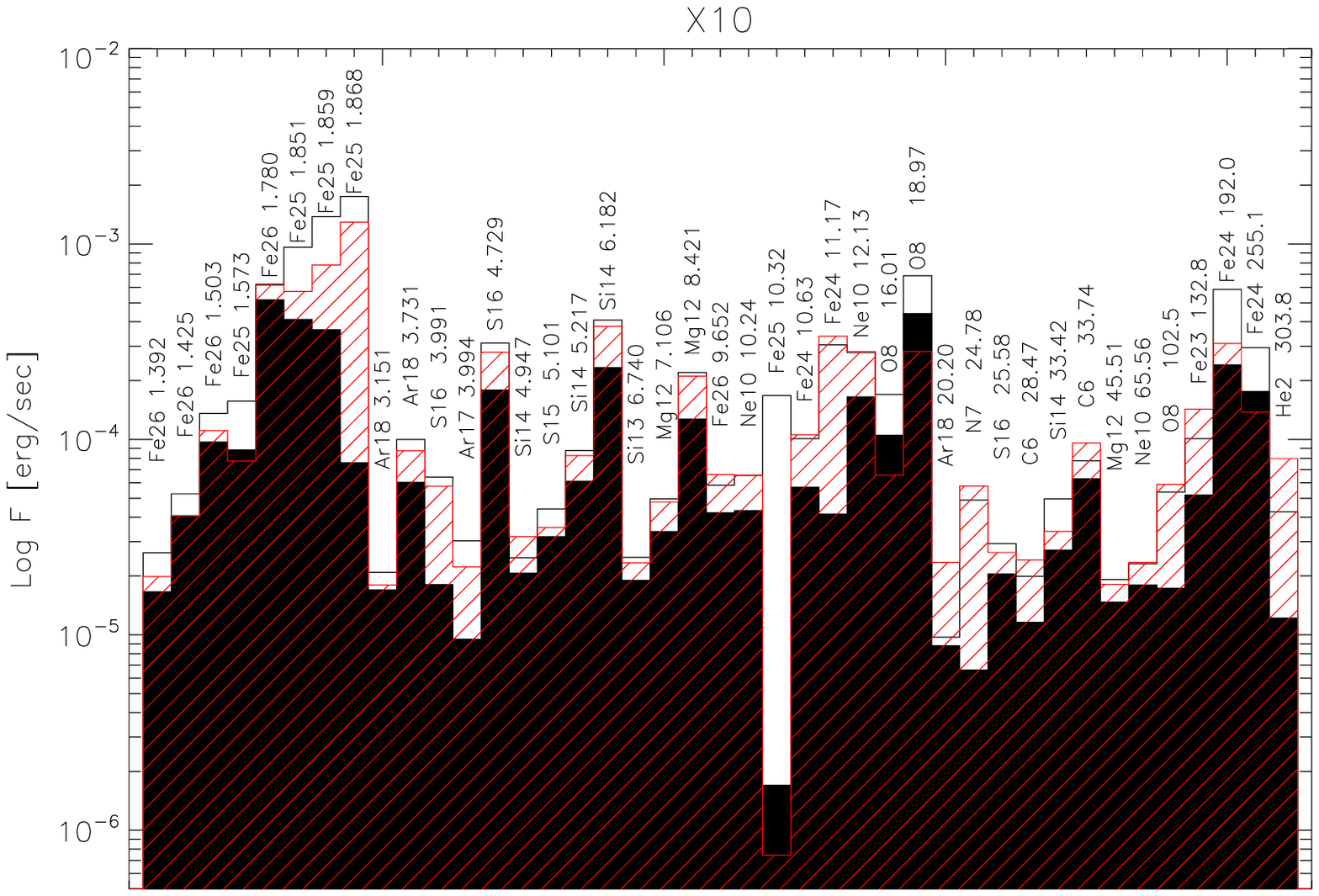}
\caption{\footnotesize X-ray slab; U$_{\rm X}$=10 (X10). The red,
  striped histograms illustrate {\sc mocassin's} results, while the
  filled and empty histograms represent the lower and higher limits of the
  range of predictions obtained by the codes in P\'equignot et al. (2001) . 
\label{f:x10}
}
\end{figure}

In Figures~\ref{f:x01}, \ref{f:x1} and \ref{f:x10} we use red
(striped) histograms to illustrate {\sc mocassin}'s results, while the filled and
empty black histograms respectively mark the lower and the higher limit of the
range of solutions obtained by the other codes in the original
exercise. All line fluxes are in units of $[erg\,s^{-1}]$ for a column
density of 10$^{16}\,cm^{-2}$, with the emissivities being summed over 4$\pi$.
In aid of future benchmarking efforts we provide in
Table~\ref{t:xlines} the individual line fluxes in $[erg\,s^{-1}]$
as predicted by this version (3.00) of {\sc mocassin}.

In spite of all the complications listed above, the agreement of {\sc
  mocassin}'s predictions with the benchmarks is reassuring and adds
  confidence to current and future astrophysical applications of the
  code. In particular, very good agreement is shown for the relative
  strengths of lines from far infrared to X-ray wavelengths. Some
  exceptions are discussed in more detail below. 

The agreement between {\sc mocassin} and the codes in P\'equignot et
al. (2001)  for the benchmark X01 is good, although the electron temperature of the slab,
listed in Table~\ref{t:xlines}, carries an isolation factor of
1.14. While this is small and not of concern {\it per se } 
it may affect the strengths of some of the emission
lines.

Figures~\ref{f:x1} and~\ref{f:x10} demonstrate a very good agreement between {\sc mocassin} and
the codes in P\'equignot et al. (2001) for most of the lines listed in
Table~\ref{t:xlines} for benchmarks X1 and X10. 
The small discrepancies with some of the H-like recombination lines (e.g. He~{\sc ii}~303.8~{\AA} and H~{\sc
  i}~1216~{\AA} in X1 and Ar~{\sc xviii}~20.20~{\AA},
N~{\sc vii}~24.78~{\AA}, C~{\sc vi}~28.47~{\AA} and 33.74~{\AA} and He~{\sc
  ii}~303.8~{\AA} in X10) are most probably due to different
extrapolation techniques of the Storey \& Hummer (1995) data to the
high temperatures of these models. 

\section{Summary}\label{s:summary}

We have presented a new version of the fully 3D
 Monte Carlo photoionization code, {\sc mocassin}. The code was
 extended to allow the modelling of plasma irradiated by a hard
 continuum spanning from radio to gamma rays. The atomic data set of
 the code was also significantly updated and it is now synchronized
 with the latest release of the {\sc chianti} database. 
The applicability and
 limitations of the new code were discussed, and the results
 of a thorough benchmarking exercise presented. 

No major problems were found by the benchmark tests, although some
minor differences have been found. We have
highlighted a number of significant improvements in the atomic
datasets available today compared to those available at the time the
original benchmark tables were compiled. We provide here updated
values and we emphasize the need of a new benchmarking exercise to be
undertaken by the plasma modelling community. 

The good performance of the {\sc mocassin} code in all benchmark
tests demonstrates that it is ready for application to real
astrophysical problems. The public version of the X-ray enhanced {\sc
  mocassin} code is available on request from the author. 

\section*{Acknowledgments}

We thank Daniel P\'{e}quignot for crucial guidance with the benchmark
comparisons. We also would like to thank Gary
Ferland for helpful discussion on solving the ionization balance
matrix for non-adjacent stages. Our thanks are also extended to Pete
Storey for help with the hydrogenic data set and dielectronic
recombination data. We thank the {\sc chianti} team for making the new
data available to us. Finally, we thank the anonymous referee for constructive
comments that helped to significantly improve the quality of this work.
JJD was supported by the Chandra X-ray Center NASA contract NAS8-39073
during the course of this research.  BE was supported by {\it Chandra}
grants GO6-7008X and GO6-7009X.


\begin{table}
\caption{The number of energy levels used for each iso-electronic
  sequence when compiling the {\sc mocassin} atomic data files.}
\begin{center}
\begin{tabular}{llll}
\noalign{\hrule}
\noalign{\smallskip}
Iso-electronic &Number &Iso-electronic &Number \\
Sequence & of levels &Sequence & of levels \\
\noalign{\hrule}
\noalign{\smallskip}
H  & 9   & Na & 3 \\
He & 17  & Mg & 5 \\
Li & 15  & Al & 12 \\
Be & 10  & Si & 27 \\
B  & 10  & P  & 5 \\
C  & 15  & S  & 5 \\
N  & 13  & Cl & 2 \\
O  & 9   & Ar & -- \\
F  & 2   & K  & 2  \\
Ne & --  & Ca & 9  \\
   &     & Sc &19 \\
\noalign{\hrule}
\end{tabular}
\end{center}
\label{tbl.levels}
\end{table}

\begin{table}
\begin{center}
\caption{Input parameters for benchmark models.\tablenotemark{a}\label{t:benchinput}}
\tablenotetext{a}{ Plane parallel
  geometry is assumed for all models. PL1: power law, slope 1.3; BPL:
  Broken Power Law (Table~\ref{t:bpl}). All abundances are given by
  number with respect to hydrogen.}
\begin{tabular}{lcccc}
\tableline
Model:             & NLR & X$^{\rm thin}_{0.1}$ & X$^{\rm thin}_{1}$ &
X$^{\rm thin}_{10}$ \\
\tableline
Ionizing Spectrum  & PL1 & BPL             & BPL             & BPL\\
$n_{\rm H}\,[$cm$^{-3}]$& 10$^4$ & 10$^5$ & 10$^5$ & 10$^5$ \\
$N{\rm (H)}^{\rm out} [$cm$^{-2}]$& 10$^{22}$ & 10$^{16}$ & 10$^{16}$
& 10$^{16}$ \\
U$_{13.6\,eV}$     & 0.01 & -- & -- & -- \\ 
U$_{100\,eV}$     & -- & 0.1 & 1 & 10 \\
He/H                   & 0.1    & 0.1   & 0.1	& 0.1	\\
C/H\,$\times$\,10$^5$  & 30.0   & 37.0	& 37.0	& 37.0	\\
N/H\,$\times$\,10$^5$  & 10.0   & 11.0	& 11.0	& 11.0	\\
O/H\,$\times$\,10$^5$  & 80.0   & 80.0	& 80.0	& 80.0	\\
Ne/H\,$\times$\,10$^5$ & 10.0   & 11.0	& 11.0	& 11.0	\\
Mg/H\,$\times$\,10$^5$ & 3.0    & 3.7	& 3.7	& 3.7	\\
Si/H\,$\times$\,10$^5$ & 3.0    & 3.5	& 3.5	& 3.5	\\
S/H\,$\times$\,10$^5$  & 1.5    & 1.6	& 1.6	& 1.6	\\
Ar/H\,$\times$\,10$^5$ &  --    & 0.37	& 0.37	& 0.37	\\
Fe/H\,$\times$\,10$^5$ & --     & 4.0   & 4.0   & 4.0   \\
\tableline\noalign{\smallskip}
\end{tabular}
\end{center}
\end{table}

\begin{table}
\begin{center}
\caption{Broken power law for optically thin X-ray models.\label{t:bpl}}
\begin{tabular}{lcccccc}
\tableline
log h$\nu$ (eV) & -4.8663 & -0.8663 & 1.6108 & 2.0000 & 5.0000 &
7.1337 \\
log F$_{\nu}$ (erg) & 1.0 & 14.5 & 12.7 & 10.6 & 7.6 & 1.0 \\ 
\tableline\noalign{\smallskip}
\end{tabular}
\end{center}
\end{table}

\begin{table}
\begin{center}
\caption{Standard Narrow Line Region (NLR). \tablenotemark{a}\label{t:nlr}}
\tablenotetext{a}{
The $Min$ and $Max$ values
  of each $Quantity$ were taken from the range of values predicted by
  the various codes listed in Table~7 of P\'{e}quignot et al. (2001).
  When {\sc mocassin}'s prediction for a given quantity lies outside
  the range $[Min,Max]$ the isolation factor is defined as the
  positive ratio of {\sc mocassin}/$Max$ if {\sc mocassin} $>$ $Max$
  or the negative ratio of $Min$/{\sc mocassin} if {\sc mocassin} $<$
  $Min$. All line intensities are given relative to H$\beta$.}
\begin{tabular}{lcccc}
\tableline
 Quantity          &  Min      &   Max    & {\sc mocassin}& isolation \\
\tableline
 H$\beta$ erg/s/cm$^2$ & 1.09    & 1.49     & 1.24     & 0 \\                   
 H$\beta$ 4861       & 1.00      & 1.00     & 1.00     & 0 \\
 He~{\sc i} 5876     & 0.100     & 0.139    & 0.111    & 0 \\
 He~{\sc ii} 4686    & 0.226     & 0.260    & 0.226    & 0.\\
 C~{\sc ii} $]$2325+ & 0.362     & 0.600    & 0.438    & 0 \\           
 CII1335             & 0.09      & 0.148    & --       & 0 \\		
 CIII$]$1907+1909    & 2.33      & 4.09     & 3.20     & 0.    \\	
 CIV1549+            & 3.36      & 4.90     & 3.84     & 0.	  \\	
 $[$NI$]$5200+5198   & 0.034     & 0.230    & 0.033    & -1.03	  \\    
 $[$NII$]$6584+6548  & 1.19      & 3.27     & 0.933    & -1.28	  \\	
 $[$NII$]$5755       & 0.018     & 0.310    & 0.013    & -1.39	  \\	
 NIII$]$1749+        & 0.177     & 0.265    & 0.157    & -1.12	  \\	
 $[$NIII$]$57.3$\mu$m& 0.042     & 0.050    & 0.041    & -1.03	  \\	
 NIV$]$1487+         & 0.203     & 0.250    & 0.343    & +1.37	  \\	 
 NV1240+             & 0.106     & 0.191    & 0.096    & -1.10	  \\	
 $[$OI$]$63.1$\mu$m  & 0.220     &  3.10    & 0.156    & -1.43	  \\	 
 $[$OI$]$6300+6363   & 1.37      & 2.17     & 1.44     & 0        \\	
 $[$OII$]$3726+3729  & 1.25      & 1.85     & 1.23     & -1.02	  \\	
 $[$OIII$]$51.8$\mu$m & 0.667     & 0.858   & 0.919    & +1.07    \\	 
 $[$OIII$]$88.3$\mu$m & 0.092     & 0.112   & 0.117    & +1.04    \\	 
 $[$OIII$]$5007+4959 & 31.2      & 34.9     & 39.3     & +1.12    \\	
 $[$OIII$]$4363      & 0.286     & 0.348    & 0.273    & -1.04    \\	 
 $[$OIV$]$25.9$\mu$m & 1.72      & 2.53     & 1.948    & 0	  \\	
 OIV$]$1403+         & 0.311     & 0.510    & 0.374    & 0	  \\	
 OV$]$1218+          & 0.120     & 0.201    & 0.148    & 0	  \\	
 OVI1034+            & 0.030     & 0.059    & 0.041    & 0	  \\	
 $[$NeII$]$12.8$\mu$m    & 0.145     & 0.832& 0.165    & 0        \\    
 $[$NeIII$]$15.5$\mu$m   & 1.38      & 2.86 & 1.44     & 0	  \\	
 $[$NeIII$]$3869+68  & 1.64      & 2.46     & 1.92     & 0	  \\	
 $[$NeIV$]$2423+     & 0.394     & 0.586    & 0.387    & -1.02    \\	 
 $[$NeV$]$3426+3346  & 0.520     & 0.670    & 0.456    & -1.14    \\	 
 $[$NeV$]$24.2$\mu$m     & 0.162     & 0.450& 0.340    & 0	  \\	
 $[$NeVI$]$7.63$\mu$m    & 0.226     & 0.288& 0.243    & 0	  \\	
 MgII~2798+          & 1.23      & 2.33     & 1.77     & 0.	  \\	
 $[$MgIV$]$4.49$\mu$m    & 0.092     & 0.116    &0.060    & -1.53 \\
 $[$MgV$]$5.61$\mu$m     & 0.100     & 0.230    &0.170    & 0	  \\
 $[$SiII$]$34.8$\mu$m    & 0.728     &  1.37    &0.664    & -1.09    \\
 SiII$]$2335+        & 0.114     & 0.257        &0.177    & 0	  \\
 SiIII$]$1892+       & 0.089     & 0.312        &0.130    & 0    \\
 SiIV1397+           & 0.078     & 0.121        &0.101    & 0    \\
 $[$SII$]$6716+6731  & 0.750     &  1.86        &1.80     & 0	  \\
 $[$SII$]$4069+4076  & 0.082     & 0.180        &0.222    & +1.23    \\
 $[$SIII$]$18.7$\mu$m    & 0.560     & 0.968    &0.284    & -1.97    \\
 $[$SIII$]$33.6$\mu$m    & 0.302     & 0.488    &0.140    & -2.16    \\
 $[$SIII$]$9532+9069 & 1.98      & 2.32         &1.30     & -1.54    \\
 $[$SIV$]$10.5$\mu$m     & 0.850     &  1.74    &1.57     & 0	  \\
T$_{inner} [$K$]$    & 16840     & 17100    & 16250    & -1.04    \\
$<T[N_{\rm H^+}N_{\rm e}]> [$K$]$ & 11970 & 12920 & 12480 & 0     \\
$<$He$^+$/He$>$/$<$H$^+$/H$>$ & 0.729 & 0.766 & 0.736 & 0         \\
\tableline\noalign{\smallskip}
\end{tabular}
\end{center}
\end{table}

\begin{table*}
\begin{center}
\caption{X-ray irradiated slabs. Emission line fluxes predicted by
  {\sc mocassin}, given in units of $[erg/s]$. \label{t:xlines}
}
\begin{tabular}{lcc|lcc|lcc}
\tableline
\multicolumn{3}{c}{X01} &\multicolumn{3}{c}{X1} &\multicolumn{3}{c}{X10} \\
Ion   & $\lambda\,[${\AA}$]$ & Flux & Ion   & $\lambda\,[${\AA}$]$ & Flux & Ion   & $\lambda\,[${\AA}$]$ & Flux \\
\tableline
Mg9   & 9.314  &  1.17E-04  & S16   & 4.792    &  7.05E-05   &Fe26 & 1.392 & 1.98E-05 \\
Ne10  & 12.13  &  1.30E-04  & S15   & 5.101    &  1.40E-04   &Fe26 & 1.425 & 4.06E-05 \\
Ne9   & 13.45  &  1.29E-04  & Sil14 & 6.182    &  2.30E-04   &Fe26 & 1.503 & 1.11E-04 \\
Ne9   & 13.55  &  1.53E-04  & Sil13 & 6.648    &  7.32E-05   &Fe25 & 1.573 & 7.75E-05 \\
Ne9   & 13.70  &  4.27E-04  & Sil13 & 6.688    &  1.08E-04   &Fe26 & 1.780 & 6.23E-04 \\
O8    & 15.18  &  1.19E-04  & Sil13 & 6.740    &  2.31E-04   &Fe25 & 1.851 & 5.72E-04 \\
O8    & 16.01  &  3.27E-04  & Mg12  & 7.106    &  4.91E-05   &Fe25 & 1.859 & 7.81E-04 \\
O8    & 18.97  &  1.84E-03  & Mg12  & 8.421    &  2.38E-04   &Fe25 & 1.868 & 1.29E-03 \\
N7    & 20.91  &  3.86E-05  & Mg11  & 9.314    &  8.98E-05   &Ar18 & 3.151 & 1.79E-05 \\
O7    & 21.60  &  4.57E-04  & Ne10  & 10.24    &  9.04E-05   &Ar18 & 3.731 & 8.75E-05 \\
O7    & 21.80  &  4.78E-04  & Ne10  & 12.13    &  4.14E-04   &S16  & 3.991 & 5.76E-05 \\
O7    & 22.10  &  1.58E-03  & Ne9   & 13.70    &  5.61E-05   &Ar17 & 3.994 & 2.23E-05 \\
N7    & 24.78  &  2.07E-04  & O8    & 15.18    &  1.15E-04   &S16  & 4.729 & 2.79E-04 \\
C6    & 28.47  &  9.70E-05  & O8    & 16.01    &  2.94E-04   &Si14 & 4.947 & 3.17E-05 \\
N6    & 29.54  &  1.24E-04  & O8    & 18.97    &  1.26E-03   &S15  & 5.101 & 3.54E-05 \\
C6    & 33.74  &  5.04E-04  & N7    & 24.78    &  9.64E-05   &Si14 & 5.217 & 8.27E-05 \\
C5    & 41.47  &  8.10E-05  & C6    & 33.74    &  1.68E-04   &Si14 & 6.182 & 3.78E-04 \\
O8    & 102.5  &  1.94E-04  & Fe19  & 101.5    &  3.23E-05   &Si13 & 6.740 & 2.34E-05 \\
C6    & 182.2  &  5.01E-05  & O8    & 102.5    &  1.05E-04   &Mg12 & 7.106 & 4.78E-05 \\
He2   & 256.0  &  2.32E-04  & Fe19  & 108.4    &  1.28E-04   &Mg12 & 8.421 & 2.10E-04 \\
Fe15  & 284.2  &  6.87E-04  & Fe22  & 101+17+36&  4.33E-04   &Fe26 & 9.652 & 6.60E-05 \\
He2   & 303.8  &  9.38E-04  & Fe20  & 121.8    &  3.03E-04   &Ne10 & 10.24 & 6.57E-05 \\
Si11  & 303.9  &  2.77E-04  & Fe21  & 128.0    &  6.14E-04   &Fe25 & 10.32 & 7.46E-07 \\
Fe13  & 316.0  &  8.33E-05  & Fe20  & 132.8    &  4.44E-04   &Fe24 & 10.63 & 1.05E-04 \\
Fe14  & 333.7  &  1.39E-04  & Fe23  & 132.8    &  3.64E-04   &Fe24 & 11.17 & 3.37E-04 \\
Fe16  & 335.4  &  1.52E-04  & Fe24  & 192.0    &  7.72E-05   &Ne10 & 12.13 & 2.81E-04 \\
Fe16  & 360.8  &  9.27E-05  & Fe24  & 255.1    &  3.90E-05   &O8   & 16.01 & 6.57E-05 \\
Mg9   & 368.1  &  5.32E-04  & He2   & 303.8    &  1.64E-04   &O8   & 18.97 & 2.81E-04 \\
Ne7   & 465.0  &  3.22E-04  & Ar16  & 389+     &  9.34E-05   &Ar18 & 20.20 & 2.34E-05 \\
Sil12 & 506+  &  3.05E-04  & S14   & 418+     &  1.58E-04   &N7   & 24.78 & 5.77E-05 \\
Si11  & 581.0  &  8.21E-05  & Si12  & 506+     &  7.30E-05   &S16  & 25.58 & 2.64E-05 \\
Mg10  & 615+23 &  1.04E-03  & Fe20  & 567.8    &  4.82E-05   &C6   & 28.47 & 2.41E-05 \\
O5    & 629.7  &  6.30E-05  & Fe20  & 721.4    &  9.57E-05   &Si14 & 33.42 & 3.38E-05 \\
Ne8   & 770+80 &  1.47E-03  & Fe22  & 845.4    &  8.52E-05   &C6   & 33.74 & 9.59E-05 \\
Ne7   &895.2   &  5.39E-05  & H1    & 1216     &  9.13E-05   &Mg12 & 45.51 & 1.81E-05 \\
O6    &1034    &  8.74E-04  & Fe21  & 1354     &  2.14E-04   &Ne10 & 65.56 & 2.34E-05 \\
H1    &1216    &  4.97E-04  & T/10$^5$K &      &   6.99       &O8 & 102.5 &  5.89E-05  \\
Fe13  & 2579   &  6.12E-06  &       &          &            &Fe23 & 132.8 &  1.42E-04  \\
Fe13  & 3388   &  5.60E-06  &       &          &            &Fe24 & 192.0 &  3.09E-04  \\
Fe14  & 5303   &  9.21E-04  &       &          &            &Fe24 & 255.1 &  1.38E-04  \\
S12   &7536    &  1.02E-04  &       &          &            &He2  & 303.8 &  7.97E-05  \\
Fe13  & 10747  &  7.91E-05  &       &          &            &T/10$^6$K&   & 1.44        \\
Fe13  & 10798  &  5.22E-05  &       &          &            &   &   &\\
Si10  & 14302  &  2.95E-04  &       &          &            &   &   &\\
T/10$^4$K  &   &  12.2      &       &          &            &   &   &\\
\tableline\noalign{\smallskip}
\end{tabular}
\end{center}
\end{table*}		   

\end{document}